\documentclass{article}
\usepackage{amssymb}
\usepackage{amsmath}

\begin{document}
\centerline{\Large \bf The general solutions of some nonlinear }
\medskip
\centerline{\Large \bf second and third order PDEs}
\medskip
\centerline{\Large \bf with constant and nonconstant parameters}

\vskip 2cm \centerline{\sc Yu. N. Kosovtsov}
\medskip
\centerline{Lviv Radio Engineering Research Institute, Ukraine}
\centerline{email: {\tt kosovtsov@escort.lviv.net}} \vskip 1cm
\begin{abstract}
Selection of 25 examples from extensive nontrivial families for
different types of nonlinear PDEs and their formal general solutions
are given. The main goal here is to show on examples the types of
solvable PDEs and what their general solutions look like.
\end{abstract}

\section{Introduction}

Nonlinear partial differential equations (PDEs) play very important
role in many fields of mathematics, physics, chemistry, and biology,
and numerous applications. Despite the fact that various methods for
solving nonlinear PDEs have been developed in 19-20 centuries \cite
{Darb1}-\cite {Zhiber}, there exists a very disadvantageous opinion
that only a very small minority of nonlinear second- and
higher-order PDEs admit general solutions in closed form (see, e.g.,
percentage of PDEs with general solutions in fundamental handbook
\cite {Pol}).

Nevertheless there exist some extensive nontrivial families for
different types of nonlinear PDEs which general solutions can be
expressed in closed form and which seemingly are not described in
literature.

In present paper, as a preliminary result, some relatively simple
examples of such PDEs of second and third order and their formal
general solutions are given. The main goal here is to show on
examples the types of solvable PDEs and what their general solutions
look like.

\section{Notations remarks}

The expression of the following type \[RootOf[F(\_Z)]\] means
\emph{any} root of the algebraic equation $F(\_Z)=0$ with respect to
indeterminate $\_Z$, and
\[X^{\frac{m}{n}}=RootOf[\_Z^{\frac{n}{m}}-X]\,.\]

For shortness \[\int^xf(t,x)dt=\{\int f(t,x)dt\}|_{t=x}\,.\] \vskip
0.5cm

\section{Second order PDEs with two independent \\variables and
constant parameters}

\vskip 1cm


\subsection {\mathversion{bold}{{\Large$\frac{\partial^2 w}{\partial t
\partial x}-(\frac{1}{w}\frac{\partial w}{\partial t}+b)\frac{\partial w}{\partial x}-\frac{c}{w}\frac{\partial w}{\partial t}-c b
= 0\,,$}}} \vskip 0.5cm

where $b,c$ are constants. \vskip 0.5cm General solution

\begin{equation}
w(t,x) = [-c\int\,\exp[-e^{bt}G(x)]\,dx+F(t)]\exp[e^{bt}G(x)]\,,
\notag
\end{equation}
where $F(t)$ and $G(x)$ are arbitrary functions. \vskip 0.5cm


\subsection {\mathversion{bold}{{\Large$\frac{\partial^2 w}{\partial t
\partial x}-\frac{1}{w}\frac{\partial w}{\partial t}\frac{\partial w}{\partial x}-\frac{c}{w}\frac{\partial w}{\partial t}-k w
= 0\,,$}}} \vskip 0.5cm

where $c,k$ are constants.\vskip 0.5cm General solution

\begin{equation}
w(t,x) =
\frac{\{-c\int\exp[x(1-kt)]G(x)\,dx+F(t)\}\exp[-x(1-kt)]}{G(x)}\,,
\notag
\end{equation}
where $F(t)$ and $G(x)$ are arbitrary functions. \vskip 0.5cm


\subsection {\mathversion{bold}{{\Large$\frac{\partial^2 w}{\partial t
\partial x}=\frac{a}{w}(\frac{\partial w}{\partial x})^2+(\frac{1}{w}\frac{\partial w}{\partial t}+b+\frac{c}{w})\frac{\partial w}{\partial x}+\frac{2c\frac{\partial w}{\partial
t}+(bw+c)^2}{4aw}\,,$}}} \vskip 0.5cm where $a,b,c$ are constants,
and $a\neq0$. \vskip 0.5cm General solution

\begin{equation}
w(t,x) =
\{-\frac{c}{2a}\int\exp[\frac{1}{2a}\int\frac{2\,dx}{t+G(x)}+bx]dx+F(t)\}\exp[-\frac{1}{2a}\int\frac{2\,dx}{t+G(x)}+bx]
\,, \notag
\end{equation}
where $F(t)$ and $G(x)$ are arbitrary functions. \vskip 0.5cm


\subsection {\mathversion{bold}{{\Large$\frac{\partial^2 w}{\partial t
\partial x}-(\frac{1}{w}\frac{\partial w}{\partial t}+b)\frac{\partial w}{\partial x}-\frac{c}{w}\frac{\partial w}{\partial t}-kw-cb
= 0\,,$}}} \vskip 0.5cm

where $b,c,k$ are constants, and $b\neq0,k\neq0$. \vskip 0.5cm
General solution

\begin{equation}
w(t,x) =
[-c\int\exp(\frac{k}{b^2}[e^{bt}G(x)+bx])dx+F(t)]\exp(-\frac{k}{b^2}[e^{bt}G(x)+bx])
\,, \notag
\end{equation}
where $F(t)$ and $G(x)$ are arbitrary functions. \vskip 0.5cm


\subsection {\mathversion{bold}{{\Large$\frac{\partial^2 w}{\partial t\partial x}-\frac{a}{w}(\frac{\partial
w}{\partial x})^2-(\frac{1}{w}\frac{\partial w}{\partial
t}+b+\frac{c}{w})\frac{\partial w}{\partial
x}\\
\\-\frac{c}{2aw}\frac{\partial w}{\partial
t}-kw-\frac{bc}{2a}-\frac{c^2}{4aw} = 0\,,$}}} \vskip 0.5cm

where $a,b,c,k$ are constants,and $a\neq0,b^2-4ka\neq0$. \vskip
0.5cm General solution

\begin{align}
&w(t,x) =\notag\\\notag\\&
-\frac{c}{2a}\{\int\exp[\frac{1}{2a}\int\frac{\exp(t\sqrt{b^2-4ak})G(x)(b+\sqrt{b^2-4ak})-\sqrt{b^2-4ak}+b}{1+\exp(t\sqrt{b^2-4ak})G(x)}dx]dx\notag\\\notag\\&
+F(t)\}
\exp[-\frac{1}{2a}\int\frac{\exp(t\sqrt{b^2-4ak})G(x)(b+\sqrt{b^2-4ak})-\sqrt{b^2-4ak}+b}{1+\exp(t\sqrt{b^2-4ak})G(x)}dx]
\,, \notag
\end{align}
where $F(t)$ and $G(x)$ are arbitrary functions. \vskip 0.5cm


\subsection {\mathversion{bold}{{\Large$\frac{ac}{w}+b+\frac{a}{w}\frac{\partial w}{\partial
x}-\frac{a^2}{w^4}(c\frac{\partial w}{\partial t}+\frac{\partial
w}{\partial t}\frac{\partial w}{\partial x}-w\frac{\partial^2
w}{\partial t\partial x})^2=0\,,$}}} \vskip 0.5cm

where $a,b,c$ are constants, and $a\neq0$. \vskip 0.5cm General
solution

\begin{align}
w(t,x) =
 \{-c\int
&\exp[-\frac{1}{4a}(-4bx+\int(t+G(x))^2dx)]dx+F(t)\}\notag\\\notag\\&\times\exp[\frac{1}{4a}(-4bx+\int(t+G(x))^2dx)]
\,, \notag
\end{align}
where $F(t)$ and $G(x)$ are arbitrary functions. \vskip 0.5cm


\subsection {\mathversion{bold}{{\Large$\frac{\partial^2 w}{\partial t\partial x}-(\frac{1}{w}\frac{\partial
w}{\partial t}+b)\frac{\partial w}{\partial
x}-\frac{g}{w}\frac{\partial w}{\partial t}\\\\-\frac{a
w^2}{g+hw+\frac{\partial w}{\partial x}}-kw-b g = 0\,,$}}} \vskip
0.5cm

where $a,b,g,h,k$ are constants. \vskip 0.5cm General solution

\begin{align}
&w(t,x) = \notag\\\notag\\&\{-g\int e^{hx}\exp\{-\int
RootOf[t-\int^{\_Z}\frac{\xi\,d\xi}{(k- h
b)\xi+b\xi^2+a}+G(x)]dx\}dx+F(t)\}\notag\\\notag\\&\times
e^{-hx}\exp\{\int RootOf[t-\int^{\_Z}\frac{\xi\,d\xi}{(k- h
b)\xi+b\xi^2+a}+G(x)]dx\} \,, \notag
\end{align}
where $F(t)$ and $G(x)$ are arbitrary functions. \vskip 0.5cm


\subsection {\mathversion{bold}{{\Large$\frac{\partial^2 w}{\partial t\partial
x}-(\frac{1}{w}\frac{\partial w}{\partial t}-m)\frac{\partial
w}{\partial
x}\\\\+\frac{kw}{a}\exp(\frac{ac}{w}+b+\frac{a}{w}\frac{\partial
w}{\partial x})\\\\-\frac{c}{w}\frac{\partial w}{\partial t}+\frac{g
w}{a}+cm = 0\,,$}}} \vskip 0.5cm

where $a,b,c,k,g,m$ are constants, and $a\neq0$. \vskip 1cm General
solution

\begin{align}
&w(t,x) =  \{-c\int \exp(\frac{bx} {a})\notag\\\notag\\&\times
\exp(-\frac{1}{a}\int RootOf[t+\int^{\_Z}\frac{d\xi}{g-bm+k
\exp(\xi)+m\xi} +G(x)]dx)dx+F(t)\}\notag\\\notag\\&\times
\exp(-\frac{bx} {a})\exp(\frac{1}{a}\int
RootOf[t+\int^{\_Z}\frac{d\xi}{g-bm+k \exp(\xi)+m\xi} +G(x)]dx) \,,
\notag
\end{align}
where $F(t)$ and $G(x)$ are arbitrary functions. \vskip 0.5cm


\subsection {\mathversion{bold}{{\Large$w(\frac{\partial w}{\partial x}+cw+b)\frac{\partial^2 w}{\partial
t\partial x}\\\\ = \frac{\partial w}{\partial t}\frac{\partial
w}{\partial x}(\frac{\partial w}{\partial x}+c w+2b)+b(c
w+b)\frac{\partial w}{\partial t}+m w^3\,,$}}} \vskip 0.5cm

where $b,c,m$ are constants. \vskip 1cm General solution

\begin{align}
w(t,x)=
-\{b\int\exp[-\int(\sqrt{G(x)+2mt}-c)dx]dx+F(t)\}\exp[\int(\sqrt{G(x)+2mt}-c)dx]
 \,, \notag
\end{align}
where $F(t)$ and $G(x)$ are arbitrary functions. \vskip 0.5cm


\subsection {\mathversion{bold}{{\Large$w(\frac{\partial w}{\partial x}+cw+b)^2\frac{\partial^2 w}{\partial
t\partial x} \\\\= \frac{\partial w}{\partial t}(\frac{\partial
w}{\partial x})^3+(2c w+3b)\frac{\partial w}{\partial
t}(\frac{\partial w}{\partial x})^2\\\\+(c w+b)(c
w+3b)\frac{\partial w}{\partial t}\frac{\partial w}{\partial
x}\\\\+b(c w+b)^2\frac{\partial w}{\partial t}+m w^4\,,$}}} \vskip
0.5cm

where $b,c,m$ are constants. \vskip 1cm General solution

\begin{align}
w =
 -\{b\int\exp[-\int((G(x)+3mt)^{\frac{1}{3}}-c)dx]dx+F(t)\}\exp[\int((G(x)+3mt)^{\frac{1}{3}}-c)dx]
 \,, \notag
\end{align}
where $F(t)$ and $G(x)$ are arbitrary functions. \vskip 0.5cm


\subsection {\mathversion{bold}{{\Large$w(\frac{\partial w}{\partial x}+cw+b)\frac{\partial^2 w}{\partial
t\partial x}\\\\ = a(\frac{\partial w}{\partial
x})^3+(\frac{\partial w}{\partial t}+(3ac+k)w+3ab)(\frac{\partial
w}{\partial x})^2\\\\+((c w+2b)\frac{\partial w}{\partial
t}+(\frac{k^2}{3a}+2kc+3ac^2)w^2\\\\+2b(k+3ac)w+3ab^2)\frac{\partial
w}{\partial x}+b(c w+b)\frac{\partial w}{\partial
t}\\\\+(\frac{k^3}{27a^2}+\frac{k^2c}{3a}+ac^3+kc^2)w^3\\\\+b(\frac{k^2}{3a}+2ck+3ac^2)w^2+b^2(3ac+k)w+ab^3\,,$}}}
\vskip 0.5cm

where $a,b,c,k$ are constants, and $a\neq0,k\neq0$. \vskip 1cm
General solution

\begin{align}
w(t,x) =
 -\{b\int&\exp[\frac{1}{3a}\int\frac{(k+3ac)\sqrt{6kt+G(x)}-9ac}{\sqrt{6kt+G(x)}-3}dx]dx+F(t)\}\notag\\\notag\\
&\times\exp[-\frac{1}{3a}\int\frac{(k+3ac)\sqrt{6kt+G(x)}-9ac}{\sqrt{6kt+G(x)}-3}dx]
 \,, \notag
\end{align}
where $F(t)$ and $G(x)$ are arbitrary functions.

\vskip 2cm
\section{Second order PDEs with two independent variables and
non-constant parameters}

\vskip 1cm


\subsection {\mathversion{bold}{{\Large$\frac{\partial^2 w}{\partial t
\partial x} =
a(\frac{\partial w}{\partial
x})^2-(\frac{2ahw}{hx+b}-cx-g)\frac{\partial w}{\partial
x}+\frac{h}{hx+b}\frac{\partial w}{\partial
t}\\\\+\frac{ah^2w^2}{(hx+b)^2}-\frac{h(cx+g)w}{hx+b}+\frac{(hx+b)(2hg-cb+chx)c}{4h^2a}\,,$}}}
\vskip 0.5cm

where $a,b,c,g,h$ are constants, and $a\neq0,h\neq0,hg-cb\neq0$.
\vskip 0.5cm General solution

\begin{equation}
w =
\frac{(hx+b)c}{2ah}\{F(t)-x+2(hg-cb)\int[1/(-c(hx+b)+\exp[\frac{(cb-hg)t}{h}]G(x))]dx\}\,,
\notag
\end{equation}
where $F(t)$ and $G(x)$ are arbitrary functions. \vskip 0.5cm


\subsection {\mathversion{bold}{{\Large$w[b(t,x)+c(t,x)w+\frac{\partial w}{\partial x}]\frac{\partial^2
w}{\partial t
\partial x}=
\frac{\partial w}{\partial t}(\frac{\partial w}{\partial
x})^2\\\\+[(c(t,x)w+2b(t,x))\frac{\partial w}{\partial
t}-\frac{\partial c(t,x)}{\partial t}w^2-w \frac{\partial
b(t,x)}{\partial t}]\frac{\partial w}{\partial
x}\\\\+b(t,x)[c(t,x)w+b(t,x)]\frac{\partial w}{\partial
t}-a(t,x)w^3\\\\-[\frac{\partial b(t,x)}{\partial
t}c(t,x)+\frac{\partial c(t,x)}{\partial
t}b(t,x)]w^2-w\frac{\partial b(t,x)}{\partial t}b(t,x)\,,$}}} \vskip
0.5cm

General solution

\begin{align}
w = &\{-\int b(t,x)\exp[-\int(-c(t,x)+\sqrt{c^2(t,x)-2\int
a(t,x)dt+G(x)})dx]dx+F(t)\}\notag\\\notag\\&
\times\exp[\int(-c(t,x)+\sqrt{c^2(t,x)-2\int a(t,x)dt+G(x)})dx]\,,
\notag
\end{align}
where $F(t)$ and $G(x)$ are arbitrary functions. \vskip 0.5cm


\subsection {\mathversion{bold}{{\Large$\frac{\partial^2 w}{\partial t
\partial x}=
\frac{a(t,x)}{w}(\frac{\partial w}{\partial
x})^2+[\frac{1}{w}\frac{\partial w}{\partial
t}+b(t,x)+\frac{c(t,x)}{w}]\frac{\partial w}{\partial
x}\\\\+\frac{c(t,x)}{2a(t,x)w}\frac{\partial w}{\partial
t}+\frac{c^2(t,x)}{4a(t,x)w}\\\\+\frac{1}{2a^2(t,x)}[a(t,x)b(t,x)c(t,x)\\\\+c(t,x)\frac{\partial
a(t,x)}{\partial t}-a(t,x)\frac{\partial c(t,x)}{\partial t}]\,,$}}}
\vskip 0.5cm where $a(t,x)\neq0,c(t,x)\neq0$. \vskip 0.5cm General
solution

\begin{equation}
w =
 -\frac{1}{2W(t,x)}[\int\frac{c(t,x)W(t,x)}{a(t,x)}dx+F(t)]\,, \notag
\end{equation}
where

\begin{align}
&W(t,x)=\exp \{ \int c(t,x) \exp(\int b(t,x)dt)/[-c(t,x)\notag
\\ \notag\\&\times\int \{[
\exp(\int b(t,x)dt)(a(t,x)c(t,x)(-c(t,x)+2b(t,x))\notag
\\ \notag\\&+2c(t,x)\frac{\partial
a(t,x)}{\partial t}-2\frac{\partial c(t,x)}{\partial
t}a(t,x)]/[c^2(t,x)]\}dt\notag
\\ \notag\\&-2c(t,x)G(x)+2a(t,x)\exp(\int b(t,x)dt)]dx\}\,, \notag
\end{align}

and $F(t)$ and $G(x)$ are arbitrary functions. \vskip 0.5cm


\subsection {\mathversion{bold}{{\Large$\frac{\partial^2 w}{\partial t
\partial x} =
[\frac{1}{w}\frac{\partial w}{\partial t}+b(t,x)]\frac{\partial
w}{\partial x}\\\\+\frac{g(t,x)}{w}\frac{\partial w}{\partial
t}+k(t,x)w\\\\+g(t,x)b(t,x)-\frac{\partial g(t,x)}{\partial
t}\,,$}}} \vskip 0.5cm

 \vskip 0.5cm
General solution
\begin{equation}
w =
 -\frac{1}{W(t,x)}[\int g(t,x)W(t,x)dx+F(t)]\,, \notag
\end{equation}
where
\begin{align}
&W(t,x)=\exp \{-\int \frac{\exp(\int
b(t,x)dt)}{G(x)(g(t,x)+1)}\notag
\\ \notag\\& \times[\{[G(x)(g(t,x)+1)+1]\notag
\\ \notag\\& \times\int k(t,x)\exp[-\int
b(t,x)dt]dt+(g(t,x)+1)\notag
\\ \notag\\& \times\{\int\frac{1}{(g(t,x)+1)^2}\frac{\partial
g(t,x)}{\partial t}\int k(t,x)\exp[-\int b(t,x)dt]dtdt\notag
\\ \notag\\&-\int\frac{k(t,x)}{g(t,x)+1}\exp[-\int
b(t,x)dt]dt\})\}]dx\}\,, \notag
\end{align}

and $F(t)$ and $G(x)$ are arbitrary functions.

\vskip 2cm
\section{Second order PDEs with four independent variables and
constant parameters} \vskip 1cm


\subsection {\mathversion{bold}{{\Large$A_1\frac{\partial^2 w}{\partial x_1
\partial x_4}+A_2\frac{\partial^2 w}{\partial x_2
\partial x_4}+A_3\frac{\partial^2 w}{\partial x_3
\partial x_4}+C_0+B_1\frac{\partial w}{\partial x_4}\\\\+C_1(A_1\frac{\partial w}{\partial x_1}+A_2\frac{\partial w}{\partial x_2}+A_3\frac{\partial w}{\partial x_3}+B_1w+B_0)\\\\+C_2(A_1\frac{\partial w}{\partial x_1}+A_2\frac{\partial w}{\partial x_2}+A_3\frac{\partial w}{\partial x_3}+B_1w+B_0)^2=0\,,$}}} \vskip 0.5cm

where $w=w(x_1,x_2,x_3,x_4)$ and $A_i,B_i,C_i$ are constants,\\ and
$A_1\neq0,C_2\neq0,4C_0C_2-C_1^2\neq0$. \vskip 0.5cm General
solution

\begin{align}
&w(x_1,x_2,x_3,x_4) =
-\frac{1}{2A_1C_2}\exp(-\frac{B_1x_1}{A_1})\int^{x_1}\exp(\frac{B_1\xi}{A_1})(2B_0C_2+C_1+\tan[\frac{1}{2}x_4\sqrt{4C_0C_2-C_1^2}\notag
\\ \notag \\&+G(\xi,(
A_2\xi+A_1 x_2-A_2
x_1),(A_3\xi+A_1x_3-A_3x_1))]\sqrt{4C_0C_2-C_1^2})d\xi\notag
\\ \notag \\&
+\exp(-\frac{B_1 \xi}{A_1})F((A_1x_2-A_2x_1),(A_1x_3-A_3x_1),x_4)\,,
\notag
\end{align}
where $F(t_1,t_2,t_3)$ and $G(t_1,t_2,t_3)$ are arbitrary functions.
\vskip 0.5cm


\subsection {\mathversion{bold}{{\Large$w(A_1\frac{\partial^2 w}{\partial x_1
\partial x_4}+A_2\frac{\partial^2 w}{\partial x_2
\partial x_4}+A_3\frac{\partial^2 w}{\partial x_3
\partial x_4})+C_0\\\\+C_1(w(A_1\frac{\partial w}{\partial x_1}+A_2\frac{\partial
w}{\partial x_2}+A_3\frac{\partial w}{\partial x_3})+B_0)
\\\\+C_2(w(A_1\frac{\partial w}{\partial x_1}+A_2\frac{\partial
w}{\partial x_2}+A_3\frac{\partial w}{\partial x_3})+B_0)^2
\\\\+(A_1\frac{\partial w}{\partial x_1}+A_2\frac{\partial w}{\partial
x_2}+A_3\frac{\partial w}{\partial x_3})\frac{\partial w}{\partial
x_4}=0\,,$}}} \vskip 0.5cm

where $w=w(x_1,x_2,x_3,x_4)$ and $A_i,B_i,C_i$ are constants,\\ and
$A_1\neq0,C_2\neq0,4C_0C_2-C_1^2\neq0$. \vskip 1cm General solution

\begin{align}
&w(x_1,x_2,x_3,x_4) =
\frac{1}{A_1C_2}\{A_1C_2(-\tan(\frac{x_4}{2}\sqrt{4C_0C_2-C_1^2})(2B_0C_2+C_1)
+\sqrt{4C_0C_2-C_1^2})\notag \\ \notag
\\&\times\int^{x_1}G(\xi,(\xi A_2+x_2A_1-x_1A_2),(\xi A_3+x_3A_1-x_1A_3))\notag \\ \notag &/
[-1+\tan(\frac{x_4}{2}\sqrt{4C_0C_2-C_1^2})G(\xi,(\xi
A_2+x_2A_1-x_1A_2),(\xi A_3+x_3A_1-x_1A_3))]d\xi \notag \\\notag \\
\notag &
+A_1C_2(2B_0C_2+C_1+\sqrt{4C_0C_2-C_1^2}\tan(\frac{x_4}{2}\sqrt{4C_0C_2-C_1^2}))
\notag \\ \notag
\\&\times\int^{x_1}1/[-1+\tan(\frac{x_4}{2}\sqrt{4C_0C_2-C_1^2})G(\xi,(\xi A_2+x_2A_1-x_1A_2),(\xi A_3+x_3A_1-x_1A_3))]d\xi
\notag \\
\notag \notag
\\&+F((x_2A_1-x_1A_2),(x_3A_1-x_1A_3),x_4)\}^{\frac{1}{2}} \,,
\notag
\end{align}
where $F(t_1,t_2,t_3)$ and $G(t_1,t_2,t_3)$ are arbitrary functions.
\vskip 0.5cm


\subsection {\mathversion{bold}{{\Large$w(A_1\frac{\partial^2 w}{\partial x_1
\partial x_4}+A_2\frac{\partial^2 w}{\partial x_2
\partial x_4}+A_3\frac{\partial^2 w}{\partial x_3
\partial x_4})+C_0\\\\+C_1\exp[C_2(w(A_1\frac{\partial w}{\partial
x_1}+A_2\frac{\partial w}{\partial x_2}+A_3\frac{\partial
w}{\partial x_3})+B_0)] \\\\+(A_1\frac{\partial w}{\partial
x_1}+A_2\frac{\partial w}{\partial x_2}+A_3\frac{\partial
w}{\partial x_3})\frac{\partial w}{\partial x_4}=0\,,$}}} \vskip
0.5cm where $w=w(x_1,x_2,x_3,x_4)$ and $A_i,B_i,C_i$ are constants,\\
and $A_1\neq0,C_0\neq0,C_2\neq0$. \vskip 0.5cm General solution

\begin{align}
&w(x_1,x_2,x_3,x_4) = \{-\frac{2}{A_1C_2}(B_0C_2x_1-x_1\ln
C_0+\int^{x_1}
\ln(\exp[C_2C_0(x_4\notag \\\notag \\
\notag &+G(\xi,(\xi A_2+x_2A_1-x_1A_2),(\xi
A_3+x_3A_1-x_1A_3)))]-C_1)d\xi
\notag \\\notag \\
\notag &-F((x_2A_1-x_1A_2),(x_3A_1-x_1A_3),x_4))\}^{\frac{1}{2}} \,,
\notag
\end{align}
where $F(t_1,t_2,t_3)$ and $G(t_1,t_2,t_3)$ are arbitrary functions.

\vskip 2cm

\section{Third order PDEs with two independent variables and
constant parameters}

\vskip 1cm


\subsection {\mathversion{bold}{{\Large$\frac{\partial^3 w}{\partial t^2
\partial x}=
(\frac{\frac{\partial^2 w}{\partial t^2}}{\frac{\partial w}{\partial
t}}+\frac{3}{2w}\frac{\partial w}{\partial t})\frac{\partial^2
w}{\partial t
\partial x}-\frac{3}{2w^2}(\frac{\partial w}{\partial t})^2\frac{\partial w}{\partial x}\,,$}}} \vskip 0.5cm

\vskip 0.5cm General solution

\begin{equation}
w(t,x) =\frac{G(x)}{[F(t)+H(x)]^2}\,, \notag
\end{equation}
where $F(t),G(x)$ and $H(x)$ are arbitrary functions. \vskip 0.5cm


\subsection {\mathversion{bold}{{\Large$\frac{\partial^3 w}{\partial t^2
\partial x}=
(\frac{\frac{\partial^2 w}{\partial t^2}}{\frac{\partial w}{\partial
t}}+\frac{2}{w}\frac{\partial w}{\partial t})\frac{\partial^2
w}{\partial t
\partial x}\\\\-\frac{aw^2\frac{\partial^2 w}{\partial t^2}}{\frac{\partial w}{\partial t}}
-\frac{2}{w^2}(\frac{\partial w}{\partial t})^2\frac{\partial
w}{\partial x} +2aw\frac{\partial w}{\partial t}\,,$}}} \vskip 0.5cm

\vskip 0.5cm General solution

\begin{equation}
w(t,x) = -\frac{\frac{d G(x)}{d x}}{F(t)+H(x)+atG(x)}\,, \notag
\end{equation}
where $F(t),G(x)$ and $H(x)$ are arbitrary functions. \vskip 0.5cm


\subsection {\mathversion{bold}{{\Large$\frac{\partial^3 w}{\partial t
\partial x^2} =
\frac{2}{\frac{\partial w}{\partial x}}\frac{\partial^2 w}{\partial
x^2}\frac{\partial^2 w}{\partial t
\partial x}+(b-\frac{1}{w}\frac{\partial w}{\partial t})\frac{\partial^2 w}{\partial
x^2}+\frac{g}{w}(\frac{\partial w}{\partial x})^2\,,$}}} where $b,g$
are constants, and $b\neq0$.\vskip 0.5cm

\vskip 0.5cm General solution

\begin{equation}
w(t,x) =H(t)\exp[b\int\frac{dx}{(b+g)x-e^{bt}G(x)+F(t)}]\,, \notag
\end{equation}
where $F(t),G(x)$ and $H(x)$ are arbitrary functions. \vskip 0.5cm


\subsection {\mathversion{bold}{{\Large$ [w(a w\frac{\partial^2
w}{\partial x^2}-g(\frac{\partial w}{\partial x})^2)][\frac{\partial
w}{\partial x}\frac{\partial^3 w}{\partial t
\partial x^2} -
2\frac{\partial^2 w}{\partial x^2}\frac{\partial^2 w}{\partial t
\partial x}]\\\\+a\frac{\partial w}{\partial x}\frac{\partial w}{\partial t}(\frac{\partial^2 w}{\partial x^2})^2w
-g\frac{\partial w}{\partial t}(\frac{\partial w}{\partial
x})^3\frac{\partial^2 w}{\partial x^2}+h(\frac{\partial w}{\partial
x})^5=0\,,$}}} where $a,g,h$ are constants, and $a\neq0$.\vskip
0.5cm

\vskip 0.5cm General solution

\begin{equation}
w(t,x)
=H(t)\exp[a\int\frac{dx}{(a-g)x+F(t)+\int\sqrt{G(x)-2aht}\,\,dx}]\,,
\notag
\end{equation}
where $F(t),G(x)$ and $H(x)$ are arbitrary functions. \vskip 0.5cm


\subsection {\mathversion{bold}{{\Large$[a w(a w\frac{\partial^2 w}{\partial x^2}-(a+b)(\frac{\partial
w}{\partial x})^2)]\\\\\times(\frac{\partial w}{\partial
x}\frac{\partial^3 w}{\partial t
\partial x^2}-
2\frac{\partial^2 w}{\partial x^2}\frac{\partial^2 w}{\partial t
\partial x})\\\\-a^2w(m w-\frac{\partial w}{\partial t})\frac{\partial w}{\partial x}(\frac{\partial^2 w}{\partial x^2})^2
 \\\\+a(a+b)(2mw-\frac{\partial w}{\partial t})(\frac{\partial
w}{\partial x})^3\frac{\partial^2 w}{\partial x^2}
\\\\+k(\frac{\partial w}{\partial x})^5=0\,,$}}} where $a,b,m,k$ are
constants, and $a\neq0,m\neq0$.\vskip 0.5cm

\vskip 0.5cm General solution

\begin{equation}
w(t,x)
=H(t)\exp\{-\int\frac{am\,\,dx}{bmx+F(t)+\int\sqrt{m[m(a+b)^2+k]+e^{2mt}G(x)}\,\,dx}\}\,,
\notag
\end{equation}
where $F(t),G(x)$ and $H(x)$ are arbitrary functions. \vskip 2cm

\section{Third order PDEs with two independent variables and
non-constant parameters} \vskip 1cm

\subsection {\mathversion{bold}{{\Large$w^2(w+x \frac{\partial w}{\partial
x})\frac{\partial^3 w}{\partial t
\partial x^2} \\\\=
w[xw\frac{\partial^2 w}{\partial x^2}+2x(\frac{\partial w}{\partial
x})^2+4w\frac{\partial w}{\partial x}]\frac{\partial^2 w}{\partial t
\partial x}\\\\+w^2(w^2+\frac{\partial w}{\partial t})\frac{\partial^2 w}{\partial x^2}
-2x(\frac{\partial w}{\partial x})^3\frac{\partial w}{\partial
t}\\\\-2w(w^2+2\frac{\partial w}{\partial t})(\frac{\partial
w}{\partial x})^2\,,$}}} \vskip 0.5cm

\vskip 0.5cm General solution

\begin{equation}
w(t,x) = \frac{\frac{d F(t)}{dt}}{G(x)-F(t)+xH(t)}\,, \notag
\end{equation}
where $F(t),G(x)$ and $H(t)$ are arbitrary functions. \vskip 0.5cm


\subsection {\mathversion{bold}{{\Large$\frac{\partial^3 w}{\partial t^2
\partial x} =
\frac{a}{t}\frac{\partial^2 w}{\partial t
\partial x}+\frac{k}{x}\frac{\partial^2 w}{\partial t^2}-\frac{ak}{tx}\frac{\partial w}{\partial
t}-\frac{b}{t^2}\frac{\partial w}{\partial x}+\frac{b k
w}{t^2x}\,,$}}} \vskip 0.5cm

\vskip 0.5cm General solution

\begin{align}
w(t,x) =
F(t)x^k&+H(x)t^{\frac{1}{2}+\frac{a}{2}+\frac{1}{2}\sqrt{1+2a+a^2-4b}}\notag\\\notag\\&+G(x)t^{\frac{1}{2}+\frac{a}{2}-\frac{1}{2}\sqrt{1+2a+a^2-4b}}\,,
\notag
\end{align}
where $F(t),G(x)$ and $H(x)$ are arbitrary functions. \vskip 0.5cm

\section{Conclusion}

The details of the method and more extensive lists of solvable PDEs
are preparing for publication. The method in principle fit for
implementation in CAS, e.g., in Maple.

\end{document}